# Negative to Positive Crossover of Magnetoresistance in Layered $WS_2$ with Ohmic Contact

Yangwei Zhang,[1, a)] Honglie Ning,[1, a)] Yanan Li,[1] Yanzhao Liu,[1] and Jian Wang[1, 2, 3, b)]

[1] *International Center for Quantum Materials, School of Physics, Peking University, Beijing 100871, China*

[2] *Collaborative Innovation Center of Quantum Matter, Beijing 100871, China*

[3] *State Key Laboratory of Low Dimensional Quantum Physics, Department of Physics, Tsinghua University, Beijing 100084, People's Republic of China*

The discovery of graphene has ignited intensive investigation on two dimensional (2D) materials. Among them, transition metal dichalcogenide (TMDC), a typical representative, attracts much attention due to the excellent performance in field effect transistor (FET) related measurements and applications. Particularly, when TMDC eventually reaches few-layer dimension, a wide range of electronic and optical properties, in striking contrast to bulk samples, are detected. In this Letter, we synthesized single crystalline WS2 nanoflakes by physical vapor deposition (PVD) method and carried out a series of transport measurements of contact resistance and magnetoresistance. Focused ion beam (FIB) technology was applied to deposit Pt electrodes on $WS_2$ flakes. Different from the electron beam lithography (EBL) fabricated electrodes, FIB-deposited leads exhibited ohmic contact, resolving the dilemma of Schottky barrier. Furthermore, a temperature-modulated negative-to-positive transition of magnetoresistance (MR) associated with a crossover of carrier type at similar temperature was demonstrated. Our work offers a pathway to optimize the contact for TMDC and reveals the magnetoresistance characteristics of $WS_2$ flakes, which may stimulate further studies on TMDC and corresponding potential electronic and optoelectronic applications.

---

[a)] The two authors contributed equally to this work.

[b)] Author to whom correspondence should be addressed. Electronic mail: jianwangphysics@pku.edu.cn.

The last decade has witnessed tremendous progress in 2D materials since graphene was discovered.1 TMDC, with the chemical formula $MX_2$ (M = Mo, W, Nb, Ta; X = S, Se, Te), has emerged as one of the most significant 2D system with potential for electronic application. Due to the layer-stacked structure with strong in-plane covalent bond and weak out-of-plane van der Waals (vdW) interaction, remarkable physical properties related to FET device were gradually reported in few-layer TMDCs, such as tunable carrier mobility, high subthreshold swing value and large on/off current ratio.[2,3,4] $WS_2$, a typical TMDC, has drawn much attention recently, taking account of its outstanding electronic properties. $WS_2$/graphene and $WS_2$/$MoS_2$ heterostructures have been used as vertical transistors.[5,6] Furthermore, temperature dependent transport results in liquid-gated and solid-gated multi-layered and single-layered $WS_2$ were also studied.[8,9,10,11] Though systematic FET related transport has been intensively carried out for $WS_2$, there still exists one hindrance, non-ohmic contact owing to Schottky barrier, preventing further precise investigation on transport properties. It has been reported in $MoS_2$ that the breakthrough of high carrier mobility is largely attributed to the achievement in eliminating contact resistance effects.[12,13] In addition, few investigations about MR have been reached on $WS_2$. Spurred by the unique traits and scarce reports, we conducted a systematic magnetotransport measurement on $WS_2$ nanoflakes.

In this work, we find that FIB-fabricated electrodes on $WS_2$ nanoflakes express ohmic contact property, eliminating the obstacle of Schottly barrier arising from EBL-deposited leads. Based on this ohmic contact, we carried out transport measurements focusing on magnetoresistance and Hall resistance as a complement of existing results. Though resistance against temperature at different magnetic fields exhibits similar behavior, we find the temperature dependent longitudinal MR changes from positive to negative in this system, accompanied with a sign change of carrier type. This anomaly can be comprehended on the basis of dual effect of forward interference model together with wave-function shrinkage model in a Mott variable range hopping (VRH) regime, highlighting the dominant influence of disorder induced by localized defects.

We synthesized both multi- and single-layer $WS_2$ samples. Multi-layer samples of 100nm thick were obtained through mechanic exfoliation method by using Scotch tape,[14] while the monolayer was grown by physical vapor deposition (PVD) method. WS$_2$ powders (99%, bought from Alfa) were put in the hot center of a horizontal quartz tube furnace (Lindberg/Blue M TF55035KC-1) and the SiO$_2$/Si substrate was 15-17cm away in the down-stream direction. The source was heated from room temperature to 1000°C in 40mins. The temperature was maintained for 30mins before cooling down naturally to ambient temperature. The pressure of the whole procedure should be roughly controlled around 460Pa (measured by ZDR-I-LED) with a stable 100% Ar gas flow rate of 130-160sccm (measured by MT-56).

We utilized scanning electron microscopy (SEM), energy dispersion X-ray spectroscopy (EDX), Raman Spectroscopy (RS) and atomic force microscopy (AFM) to characterize the size and thickness of the samples. Samples shown in Fig 1a inset is the SEM image of thin WS$_2$ flakes with the size around 20μm, large enough for fabrication by both FIB and EBL. W and S peaks can also be easily distinguished by EDX. Three extremely thin samples were examined by RS, and the raw curves are shown in Fig 1(b). We can see a slope of the whole curve, which is a unique property in monolayer WS$_2$.[19] After subtracting the tilt background of the spectroscopy and fitting the peaks in Lorentz mode, the average peak locations for $E_{2g}^1$ and $A_{1g}$ modes at 349.7 cm$^{-1}$ and 418.6cm$^{-1}$ with the peak intensity ratio of 0.504 were finally identified, which are all in agreement with the monolayer WS$_2$ results reported.[19] Fig 1(b) is the corresponding height profile of a flake shown in the AFM image (inset of Fig 1(b)), along the white line between the two blue crosses. A height discrepancy of approximately 0.7nm can be extracted, providing a strong evidence of single-layer WS$_2$. Therefore, we can draw a substantiated conclusion that the vapor deposition method manage to synthesize large single-layer samples.

Our six-terminal device was patterned on multi- and single-layer samples with the aid of both EBL and FIB. The transport experiment was conducted in Quantum Design Physical Property Measurement System (PPMS-16T), with field sweeping from 0T up to 15T and temperature from 300K down to 2K. By depositing electrical contacts on WS$_2$ by standard EBL method with thermal evaporation and lift-off of Ti/Au (5/50 nm), we find large non-linear contact resistance as expectation. Owing to the discrepancy of work function between the sample and the electrodes, Schottky barrier can hardly been excluded. Though Ti is already the most ideal contact metal,[14] the Schottky barrier is still not negligible in this situation. Accordingly, we need to find other methods to solve this long-standing problem.

FIB deposition is also a powerful method to make nanoscale electrodes,[15, 16, 17, 18] though the fabrication process may etch the sample and introduce impurity slightly. We fabricated Pt electrodes on $WS_2$ nanoflakes by using FIB. Surprisingly, the source-drain current exhibits a linear appearance over bias voltage not only in 100nm multilayer samples in a large temperature scale (Fig 2(a)), but also in monolayer samples (Fig. 2(b)). In sum, the contact is evidently ohmic for all temperature scale and changes little in high temperature regime. Remarkably, after subtracting the resistance of sample at 200 K, the contact resistance can be extracted, which is 980Ωμm. Thus the contact resistance at room temperature in our situation should be similar to the previously reported optimal contact values.[4, 13] The etching effect in FIB process may improve the quality of contacts. Furthermore, in terms of the previous paper which claimed Cl was a candidate to eliminate the influence of Schottky barrier and significantly reduce the contact resistance,[19] the slight Pt doing from FIB may also contribute to the observed ohmic behavior. This point is further substantiated by EDX results, in which Pt peak is clearly resolved together with W and S peaks.

Then we conducted a temperature dependent electric transport measurement on $WS_2$ nanoflakes. The whole resistance against temperature curve shows typical semiconducting behavior, with the resistance rocketing to a huge value when temperature decreases in low temperature regime. Besides, resistance as a function of temperature at the fields of 5T, 10T and 14T was also measured, resembling the resistance versus temperature behavior without field (Fig. 3(a)).

The behavior can be described in the frameworks of Mott VRH model when the temperature is down to 2K,[9, 20] which describes low temperature conductance in strongly disordered systems with localized charge-carrier states, especially in low dimensional system. The conductance satisfies the equation $\sigma = \sigma_0 exp(-(T_0/T)^{1/(1+d)})$, where $T_0$ is the hopping parameter and $d$ is the dimension of the sample.[9, 20] We fit the low temperature dependent resistance curve via different exponent referring to diverse dimension in the inset of Fig 3(a). $d = 2$ gives the best fitting, reflecting the intrinsic 2D attributes of WS$_2$ even for multilayered sample. Besides, $T_0 = 8754K$ is extracted simultaneously, from which we can eventually estimate the localization length of defects, $\xi_{loc} = (13.8/(Dk_BT_0))^{1/2}$,[9, 10] where $D$ is the density of states. Here, we use the theoretical value $D = 4\pi g_v m^*/h^2 = 2.85 \times 10^{-14} eV^{-1} cm^{-2}$, where $m^* = 0.34 m_0$ represents the electron effective mass and $g_v = 2$ stands for the valley degeneracy.[9, 21] We thus figured out the localization length $\xi_{loc} = 1.8 nm$, in good agreement with the recent reports in few-layer samples.[9, 11] The small $\xi_{loc}$ validates the existence of localized impurities in our samples which may influence the transport measurements.

Then we measured the longitudinal MR of WS$_2$ nanoflakes, defined as R(H)/R(0). Fig. 3(b) is MR with field sweeping up at several selected temperatures. The largest change of MR is approximately 4% without saturation when field goes up to 15T at 2K, as shown in Fig 3(b). The non-saturating MR has also been observed in its analogue WTe$_2$ recently.[22, 23] Remarkably, a negative to positive crossover of MR together with carrier type can be identified: when temperature is high, the MR has a negative quasi-linear slope; when temperature is from 80K down to 2K, the MR increases with the increment of field, with a quasi-linear shape (Fig 3(c)). For normal insulators or metals, the solution of Boltzmann equation yields an MR proportional to B$^2$ and usually saturates at high field. However, what we observe is a quasi-linear curve without saturation. Fig. 3(d) illustrates the temperature dependence of the slope, from which the crossover can be resolved at around 90K. Intriguingly, the absolute value of the slope diminishes with the increase of temperature in both positive and negative regions. Hall measurements were also performed at different temperatures to help understanding this phenomenon. Standard linear signals were observed, from which the carrier density was extracted and presented in Fig 3(d) as well. Negative carrier density indicates the electron-like Hall signal. It is astonishing that a crossover of carrier type simultaneously occurs at around 90K. This phenomenon indicates that when the carrier is hole-like in high temperature regime, MR exhibits a negative slope; while for electrons dominate in transport in low temperature, a positive slope of MR is observed.

Aimed at the negative to positive MR crossover, we recall three sorts of theories in semiconductors. The first is weak localization and weak anti-localization crossover, which was recently discovered in WS$_2$'s analogue WSe$_2$.[24] However, the MR is far from linearity and saturates when B > 1T. The second is magnetic polaron model. Magnetic field decreases the binding energy of the local moments introduced by impurity, giving rise to the increment of the localized length and hence negative MR.[25, 26, 27] Nevertheless, what makes this explanation invalid is the temperature dependence contrary to our results and the controversy of existence of localized magnetic order. The third possibility emphasizes the scattering in the highly disordered crystal, combining forward interference model with wave-function shrinkage model.[28, 29, 30] Forward interference model, considering the interference among different hopping paths between hopping sites, provides a negative MR term proportional to B and $T^{-3/8}$, while wave-function shrinkage model, taking account of contraction of wave-function at impurity centers in magnetic field, gives a positive term proportional to $B^2$ and $T^{-3/4}$.[28, 30] We managed to use the mentioned relation to fit the slope against temperature data well, especially in positive slope regime. In this way, we can phenomenally understand the temperature dependent behavior. When the temperature is high, forward interference model dominates and yields a negative linear MR; on the other hand, when the temperature goes lower, along with the variation of carrier type, the contraction of wave function at impurity centers dramatically reduces the localized length and the hopping probability. The wave-function shrinkage dominates, thus the system gives a positive MR. Together with Mott VRH model extracted by the temperature dependence of resistance, it is convincing to contribute the crossover of MR to this interpretation, revealing the disordered property of WS$_2$ and demonstrating the dominance of localized defects in the transport measurements.

The only enigma is the quasi-linear behavior of positive MR in contrast to the theory anticipation $B^2$. However, because of our small change of MR even in high field, it is highly possible that the MR is parabolic with very small curvature. On the other hand, Littlewood random resistor network (RRN) model,[31, 32] simulating the inhomogeneity by an N×M resistor network and giving a quasi-linear MR with adequate parameters, may explain the observed quasi-linear MR as well. Thus, the whole picture may be basically understood combined with this interpretation, again highlighting the roles of localized defects.

In summary, we utilized a simple PVD method to obtain large-scale and large-size $WS_2$ monolayer crystal. From SEM, EDX, RS and AFM, we certified the high quality of the samples. Then we used FIB method to deposit the electrodes, solving the Schottky barrier problem of the device with the leads patterned by EBL. Resistance versus temperature exhibits the Mott VRH behavior in $WS_2$ flakes. The small localization length was calculated, demonstrating the existence of strong-localized defects in $WS_2$ flake. In the end we measured MR at various temperatures and a non-saturating MR with a transition from negative to positive slope was observed associated with a crossover of carrier type at similar temperature. This intriguing behavior can be explained by the dual effect of forward interference model and wave-function shrinkage model in Mott VRH regime, substantiating the strong disorder and localized defects in $WS_2$ nanoflakes. Our measurements, along with our analysis of the data, renew the heated discussion about TMDC transport measurements and offer new strategy and perspective to explore the TMDC absorbing magnetic attributes.

We acknowledge Zhimin Liao and Huading Song for helpful discussions. This work was supported by National Basic Research Program of China grants 2013CB934600, 2012CB921300, National Natural Science Foundation of China grants 11222434, 11174007, and the Research Fund for the Doctoral Program of Higher Education (RFDP) of China.

1. K. S. Novoselov, A. K Geim, S. V. Morozov, D. Jiang, Y. Zhang, S. V. Dubonos, I. V. Grigorieva, and A. A. Firsov, Science **306** (5696), 666 (2004).
2. B. Radisavljevic, A. Radenovic, J. Brivio, V. Giacometti, and A. Kis, Nat. Nanotechnol. **6** (3),147 (2011).
3. B. Radisavljevic, and A. Kis, Nat. Mater. **12** (9), 815 (2013).
4. R. Kappera, D. Voiry, S. E. Yalcin, B. Branch, G. Gupta, A. D. Mohite, and M. Chhowalla, Nat. Mater. **13** (12), 1128 (2014).
5. K. F. Mak, C. G. Lee, J. Hone, J. Shan, and T. F. Heinz, Phys. Rev. Lett. **105** (13), 136805 (2010).
6. T. Georgiou, R. Jalil, B. D Belle, L. Britnell, R. V. Gorbachev, S. V. Morozov, Y. J. Kim, A. Gholinia, S. J. Haigh, and O. Makarovsky, Nat. Nanotechnol. **8** (2), 100 (2013).

7. X. P. Hong, J. W. Kim, S. F. Shi, Y. Zhang, C. H. Jin, Y. H. Sun, S. Tongay, J. Q. Wu, Y. F. Zhang, and F. Wang, Nat. Nanotechnol. **9**, 682 (2014).

8. W. S. Hwang, M. Remskar, R. Yan, V. Protasenko, K. Tahy, S. D. Chae, P. Zhao, A. Konar, H. L. Xing, A. Seabaugh and D. Jena, Appl. Phys. Lett. **101** (1), 013107 (2012).

9. D. Ovchinnikov, A. Allain, Y. S. Huang, D. Dumcenco, and A. Kis, ACS Nano **8** (8), 8174 (2014).

10. D. Braga, I. G. Lezama, H. Berger, and A. F. Morpurgo, Nano Lett. **12** (10), 5218 (2012).

11. F. Withers, T. H. Bointon, D. C. Hudson, M. F. Craciun, and S. Russo, Sci. Rep. **4**, 4967 (2014).

12. S. Das, H. Y. Chen, A. V. Penumatcha, and J. Appenzeller, Nano Lett. **13** (1), 100 (2012).

13. L. M. Yang, K. Majumdar, H. Liu, Y. C. Du, H. Wu, M. Hatzistergos, P. Y. Hung, R. Tieckelmann, W. Tsai, C. Hobbs, and P. D. Ye, Nano Lett. **14** (11), 6275 (2014).

14. H. L. Zeng, G. B. Liu, J. F. Dai, Y. J. Yan, B. R. Zhu, R. C. He, L. Xie, S. J. Xu, X. H. Chen, W. Yao, and S. D. Cui, Sci. Rep. **3**, 1608 (2013).

15. J. Wang, M. Singh, M. L. Tian, N. Kumar, B. Z. Liu, C. T. Shi, J. K. Jain, N. Samarth, T. E. Mallouk, and M. H. W. Chan, Nat. Phys **6**, 389 - 394 (2010).

16. M. L. Tian, J. Wang, W. Ning, T. E. Mallouk, and M. H. W. Chan, Nano Lett., **15** (3), 1487 (2015).

17. M. L. Tian, J. Wang, Q. Zhang, N. Kumar, T. E. Mallouk, and M. H. W. Chan, Nano Lett., **9** (9), 3196 (2009).

18. J. Wang, Y. Sun, M. L. Tian, B. Z. Liu, M. Singh, and M. H. W. Chan, Phys. Rev. B **86**, 035439 (2012).

19. A. Berkdemir, H. R. Gutiérrez, A. R. Botello-Méndez, N. Perea-López, A. L. Elías, C. I. Chia, B. Wang, V. H. Crespi, , F. López-Urías, J. C. Charlier, H. Terrones, and M. Terrones, Sci. Rep. **3**, 1755 (2013).

20. F. W. Van Keuls, X. L. Hu, H. W. Jiang, and A. J. Dahm, Phys. Rev. B **56** (3), 1161 (1997).

21. S. K. Kim, A. Konar, W. S. Hwang, J. H. Lee, J. Lee, J. Y. Yang, C. H. Jung, H. S. Kim, J. B. Yoo, J. Y. Choi, Y. W. Jin, S. Y. Lee, D. Jena, W. Choi, and K. Kim, Nat. Commun. **3**, 1011 (2012).

22. M. N. Ali, J. Xiong, S. Flynn, J. Tao, Q. D. Gibson, L. M. Schoop, T. Liang, N. Haldolaarachchige, M. Hirschberger, N. P. Ong and R. J. Cava, Nature **514** (7521), 205 (2014).

23. Y. F. Zhao, H. W. Liu, J. Q. Yan, W. An, J. Liu, X. Zhang, H. Jiang, Y. Liu, H. Jiang, Q. Li, Y. Wang, X. Z. Li, D. Mandrus, X. C. Xie, M. H. Pan, and J. Wang, Phys. Rev. B **92**, 041104(R)

24. H. T. Yuan, M. S. Bahramy, K. Morimoto1, S. F. Wu, K. Nomura, B. J. Yang, H. Shimotani, R. Suzuki, M. Toh, C. Kloc, X. D. Xu, R. Arita, N. Nagaosa, and Y. Iwasa, Nat. Phys. **9** (9), 563 (2013).

25. Y. B. Zhou, B. H. Han, Z. M. Liao, H. C. Wu, and D. P. Yu, Appl. Phys. Lett. **98**, 222502 (2011).

26. P. A. Bobbert, T. D. Nguyen, F. W. A. Van Oost, B. Koopmans, and M. Wohlgenannt, Phys. Rev. Lett. **99** (21), 216801 (2007).

27. T. G. Rappoport, B. Uchoa, and A. H. Castro Neto. Phys. Rev. B **80** (24), 245408 (2009).

28. H. B. Gu, J. Guo, R. Sadu, Y. D. Huang, N. Haldolaarachchige, D. Chen, D. P. Young, S. Y. Wei, and Z. H. Guo, Appl. Phys. Lett. **102** (21), 212403 (2013).

29. A. Sybous, A. El Kaaouachi, A. Narjis, L. Limouny, S. Dlimi, and G. Biskupski, J. of Mod. Phys. **3**, 521 (2012).

30. T. I. Su, C. R. Wang, S. T. Lin, and R. Rosenbaum. Phys. Rev. B **66** (5), 054438 (2002).

31. J. S. Hu, and T. F. Rosenbaum, Nat. Mater. **7** (9), 697 (2008).

32. M. M. Parish, and P. B. Littlewood, Nature **426** (6963), 162 (2003).

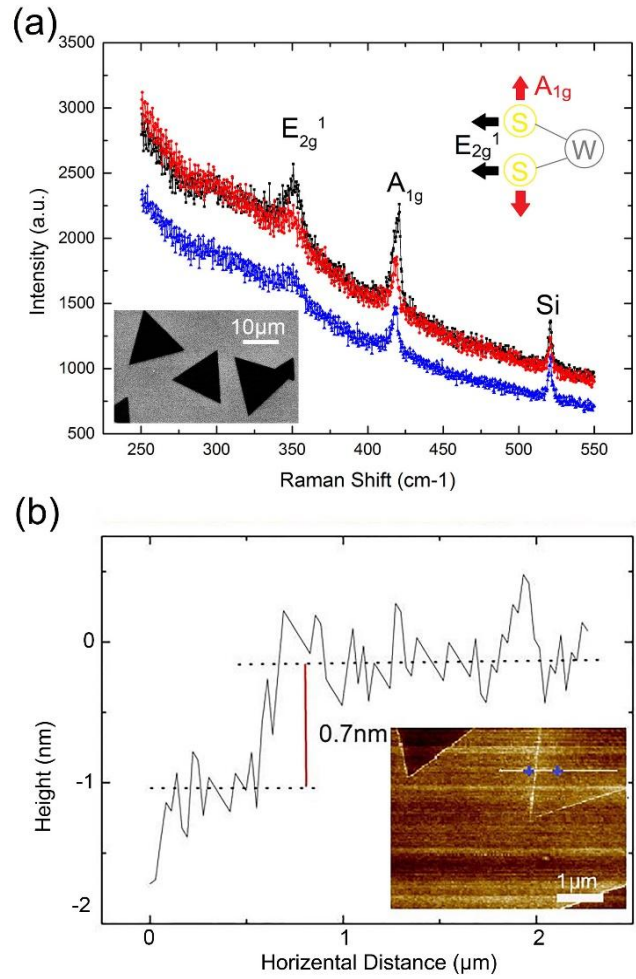

FIG. 1. Characterization of monolayer $WS_2$. (a) Raman Spectroscopy of three nanoflakes. Si signal is from the $SiO_2$ substrate. The upper inset is the schematic diagram of two first order oscillation modes and the lower inset presents a typical SEM image of the thin flakes. (b) Corresponding height profile cutting the edge of a flake shown in the inset (AFM image), along the white line between the two blue crosses, from which an average height difference of 0.7nm can be extracted.

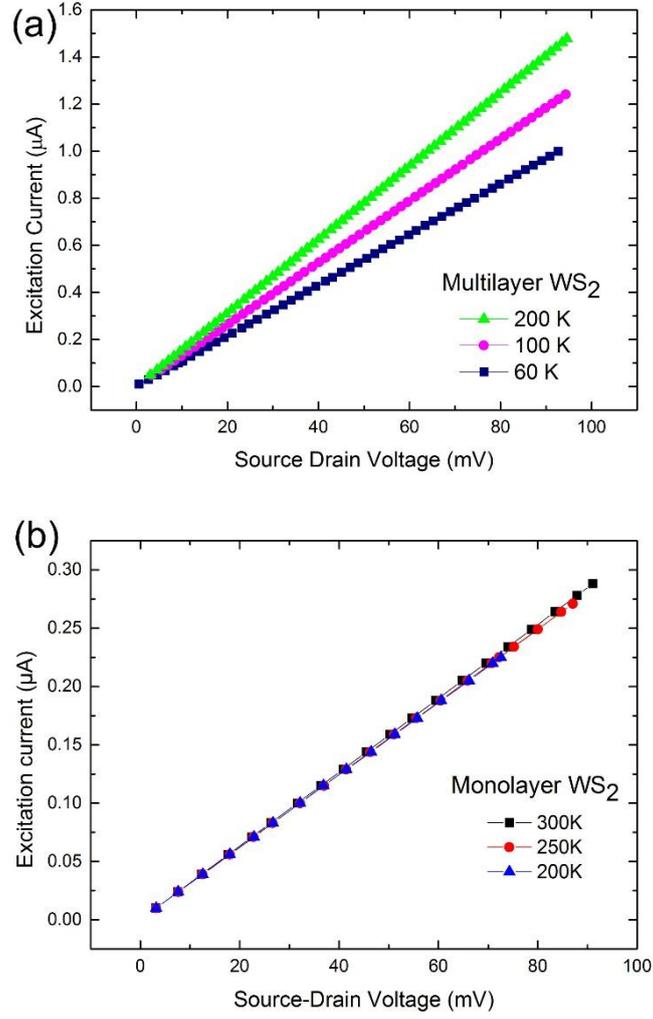

FIG. 2. Contact Resistance at different temperatures. (a) Excitation current versus source-drain voltage exhibiting ohmic behavior between the two FIB-fabricated voltage leads in a multilayer $WS_2$ flake. (b) Excitation current against source-drain voltage showing linear behavior in a monolayer $WS_2$ with FIB-fabricated electrodes.

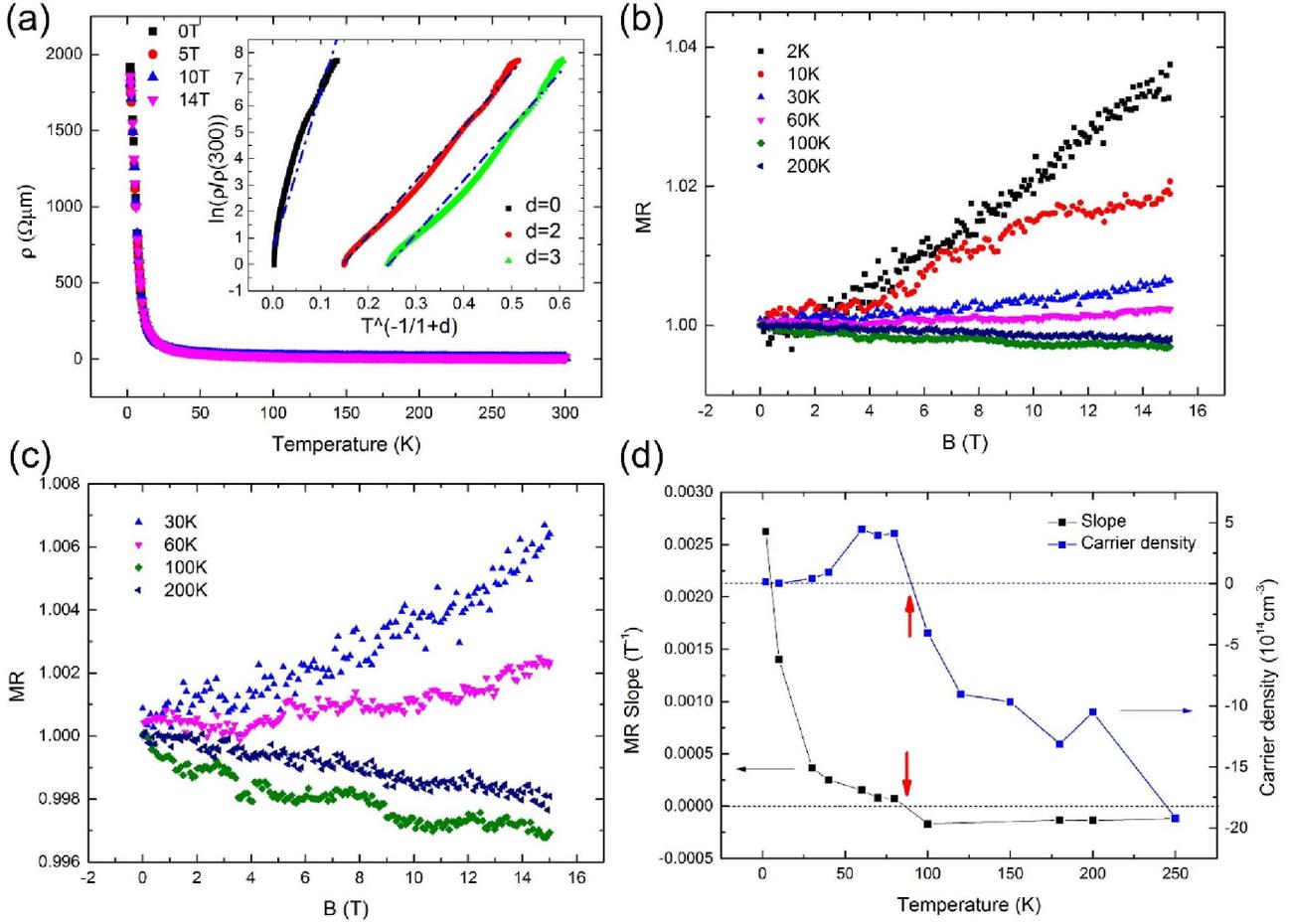

FIG. 3. Temperature dependent resistance at various magnetic fields and magnetoresistance at different temperatures. (a) Full range plot of resistivity against temperature at 0T, 5T, 10T and 15T. The temperature goes from 300K to 2K. Inset: Fitting with Mott VRH formula. Apparently d = 2 fits best. (b) MR at 6 selected typical temperatures. (c) MR at 30K, 60K, 100K and 200K, showing the crossover of quasi-linear negative to positive MR behavior. (d) Plots for slope of MR (black dots) and carrier density (blue dots) extracted from Hall measurements as a function of temperature. Carrier density above (below) zero indicates the carrier is hole (electron). The dashed black (blue) line denotes the slope of MR (carrier density) is zero. The two red arrows suggest the two crossovers of MR and carrier type.